\documentclass[epj]{svjour}
\usepackage{graphics}
\usepackage{graphicx}
\usepackage{epsfig}
\usepackage{amsmath}
\usepackage{amssymb}
\usepackage{amsfonts}
\usepackage{color}
\usepackage[latin1]{inputenc}

\begin{document}

\title{Why, when, and how fast innovations are adopted}

\author{S. Gonçalves$^1$ \and M.F. Laguna$^2$ \and J. R. Iglesias$^{1,3}$}
\mail{sgonc@if.ufrgs.br}
\institute{
Instituto de Física, Universidade Federal do Rio Grande do Sul, 
Caixa Postal 15051, 91501-970 Porto Alegre, RS, Brazil 
\and CONICET - Centro Atómico Bariloche, 8400
S. C. de Bariloche, Río Negro, Argentina. \and Programa de
Pós-Graduação em Economia Aplicada, UFRGS, and Instituto
Nacional de Ciência e Tecnologia de Sistemas Complexos, Av. João
Pessoa 52, 90040-000 Porto Alegre, RS, Brazil.}

\date{Published:6 June 2012}

\abstract{
When the full stock of a new product is quickly sold in a
few days or weeks, one has the impression that new technologies
develop and conquer the market in a very easy way. This may be true
for some new technologies, for example the cell phone, but not for
others, like the blue-ray. Novelty, usefulness, advertising, price,
and fashion are the driving forces behind the adoption of a new
product. But, what are the key factors that lead to adopt a new
technology?  In this paper we propose and investigate a simple model
for the adoption of an innovation which depends mainly on three
elements: the appeal of the novelty, the inertia or resistance to
adopt it, and the interaction with other agents.  Social interactions
are taken into account in two ways: by imitation and by
differentiation, i.e., some agents will be inclined to adopt an
innovation if many people do the same, but other will act in the
opposite direction, trying to differentiate from the ``herd''. We
determine the conditions for a successful implantation of the new
technology, by considering the strength of advertising and the effect
of social interactions. We find a balance between the advertising and
the number of anti-herding agents that may block the adoption of a new
product. We also compare the effect of social interactions,
when agents take into account the behavior of the whole society
or just a part of it.
In a nutshell, the present model reproduces
qualitatively the available data on adoption of innovation.
}


\maketitle

\section{Introduction}
\label{intro}
\begin{figure*}[htp]
\centering \includegraphics[width=12cm,clip=true]{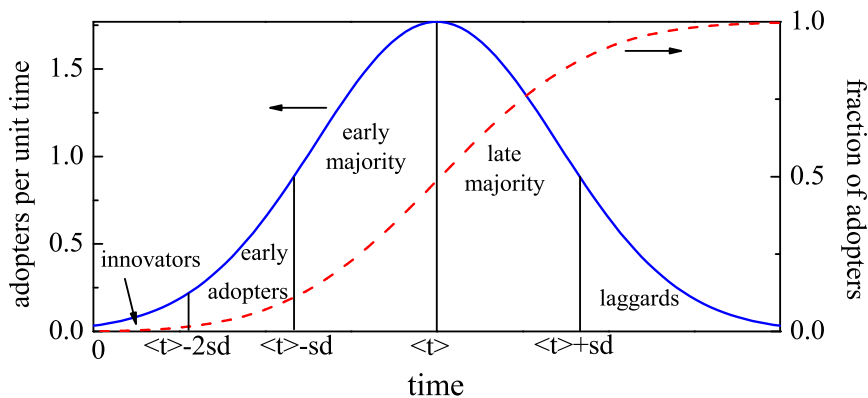}
    \caption{(color online) Dynamics of the adoption of a new
technology, according to Rogers~\cite{Rogers}. The extreme left of the
Gaussian curve (full line, in blue) correspond to the innovators,
while the extreme right to those who resist to adopt. Integrating this
Gaussian curve we obtain the total number of adopters as a function of
time that follows an error function (dashed line, in red).}
    \label{Gauss}
\end{figure*}

What are the factors that determine the adoption by the society of a
new technology? The diffusion of technology is an old and classic
problem. The usual approach is the one of Everett Rogers~\cite{Rogers}
and, according to him, people in a society are Gaussian distributed
over the time they take to attach to the novelty.  On the left side of
the distribution are the innovators, the very first ones to adopt the
new technology ---even when it has not proven to be useful or
reliable---, followed by the early adopters.  At the rightmost part
of the distribution are located the ``laggards''
who accept the innovation reluctantly, only when
the rest of the society have already changed to it.
The majority of the population fits in the middle of
the distribution, between an standard deviation of the
mean. Integration of the Gaussian gives the total number of adopters
as a function of time. A stylized representation of Rogers' ideas is
represented in Figure~\ref{Gauss}. Empirical results of the consumers
behavior of several technologies all along the 20$^{th}$ century appeared
in several sources, including an article of Brimelow in
Forbes~\cite{Forbes}, a New York Times article~\cite{NYT} and several
blogs~\cite{blog1,blog2}. All the curves exhibit a qualitative
behavior similar to the one proposed by Rogers, with the exception
that in some cases fully adoption is not attained.
Indeed, some innovations were adopted by practically the totality of
the society, like the TV or the refrigerator, while others,
like the dishwasher, were adopted by a fraction of the
households. Also, the time scales are different for different
technologies. Moreover, and as the title of Ref.~\cite{NYT} remarks:
innovations are adopted at faster speed today. Examples of this are
the following: it was necessary almost forty years (with World War II
in between) for the totality of the households to have a refrigerator,
thirty years for $90 \%$ to have a microwave oven, but only fifteen
years for almost $90 \%$ to have a cell phone. There is a tendency to
faster adoptions in recent times, even if the rapid grow of the B\&W
TV was omitted in the graphic of Ref.~\cite{NYT} (as discussed in
Ref.~\cite{blog2}).

The rapid spread of technological novelties is probably a result of
massive advertising but also of some kind of fashion and herding
effects.  As an example, many homes kept the same phone set or
refrigerator for many years, while nowadays it is usual to change the
cell phone apparatus every one or two years just because of small
changes in the functionality upgrade. We should also mention that in
some cases the old technology simply disappears and there is no
possibility of buying products with the old technology. For instance,
electronic injection in automobiles is the standard and cars are no
longer made with carburetor.  We identify three fundamental external
factors in favor of a new technology: utility, advertising pressure,
and price. The factors against changing technology are many, but we
can mention the facility of keeping a technology which people have a
good know-how, a comparable or even better quality (compare the
celluloid film with the CCD of the early digital cameras). The lower
price is also a relevant factor.  However, there is an important
factor that have been neglected in classic microeconomics
theory~\cite{masco}: the social influence.  Collective effects induced
by positive (or negative) externalities can be determinant on the
decision of buying a good~\cite{nuttall}. Understanding these collective effects is
therefore of primary importance~\cite{vicky,mirta1}.  Social
interactions are responsible of the timing in
epidemics~\cite{kup-abram}, opinion changes~\cite{castellano,us,galam},
outbreaks of social and political unrest~\cite{galam,spain}, and the
willingness to buy a given good~\cite{mirta2}. Using similar methods other 
authors studied the fashion market~\cite{galam2005} or
the dynamics of wine purchasing~\cite{galam2011}. The same kind of
social interactions is relevant for the spread of technological
innovations. A detailed discussion of social interaction effects may be found in
Refs.~\cite{vicky,castellano,galam}

Here we describe an artificial society where the external factors in
favor of the technological innovation (utility, advertising, and even
price) are reduced to a single parameter, that we call ``$A$'' and
assume as objective, i.e. the same for all the agents. This parameter
is equivalent in physics to an external field. On the other hand, the
resistance to change to a new paradigm, being because of stubbornness,
price or familiarity with the old technology, is condensed in the
idiosyncratic parameter, $u_i$, which is different for each member $i$
of the society. Finally we introduce a social interaction term that
may act reinforcing the change of technology, or going contrary to it,
if the agent resist to follow the herd. These last agents will be
called {\it contrarians}, as they will have the tendency to
act against the majority. The expression {\it contrarian} has been 
introduced by Galam~\cite{galam} but here we will use it with a slightly
different meaning, as it will be discussed below.

The article is organized as follows: In the next section we describe
the model, while in Section~\ref{results} we present the results,
first without the presence of contrarians, and then with a given
percentage of them, distinguishing the case where the agents interact
with the society as a whole or with a reduced group.  Finally, in the
last section we discuss the results and present our conclusions,
comparing with the data exhibited in Fig.~\ref{Gauss}.

\section{Model}
\label{model}
We present a model aim at describing the evolution and dynamics of
adoption of a new product, implying a new technology, after it is
placed in the market.  While consumers can either adopt (buy) or not
the novelty, our purpose is to mimic, as good as possible (by
comparison with available data) the consumers behavior at large, using
the minimal or essential ingredients.  The new product can exhibit a
completely new technological content, as the personal computer for
example, or it may imply an improvement over an existing one, like the
new models of smart phones. We assume there is a tendency for the
consumer to attach to the old technology for many reasons: to save
money, resistance to learn, etc. On the other hand there are
advantages in the new product, and the manufacturer tries to emphasize
these advantages by means of advertising.  Besides, for anyone who has
witnessed the teenagers filling the book-stores to grasp the last
Harry Potter or Twilight book, or the long lines at the computer
stores to buy the new i-Pad, it should be clear that the ``imitation''
factor has an important role too.

Rogers~\cite{Rogers} calls ``innovators'' or early adopters the people
who gather at the shop entrance the launch day: they are not sure
whether the new product is valuable, but they want it because it is
new and/or because her/his friend has one.  Regarding social
interactions, other factors come to play, as in the professional
environment, where it is important to share similar technology (smart
phones, social networks).  Other elements are not less important and
act as external pressure, like the widespread use of credit cards or
e-forms for bank accounting, airplane or hotel reservations, and tax
declarations.  The present model tries to include all these factors,
plus an additional one: the existence of people acting against
fashion, or against the majority. These are the individuals which
resist to use cell phones, microwave ovens, or make Internet shopping.

The model consider a set of $N$ interacting agents, each one
characterized by two parameters. The first one is the idiosyncratic
resistance to change represented by $u_i$ ($i$ labels the agent),
which is uniformly distributed among the members and assumes values at
random in the $[0,1]$ interval. The second parameter, $J_i$, is the
mean field coupling with the other agents. It can take positive or
negative values, the latter case representing contrarian agents. Its
effect is to weight the interaction with the people that have already
adopted the new technology. Both parameters remain fixed during the
adoption process. The marketing force or advertising effect is
represented by the parameter $A$ (the same for all members) which acts
as an external field.

The dynamics of the system is as follows: each individual $i$ checks
the value of the pay-off of adopting the innovation. The pay-off is
defined as:
\begin{equation} P = A - u_i + J_i n
\label{eq:payoff}
\end{equation}
where $n$ is the fraction of adopters of the new technology
($n=\frac{N_{adop}}{N}$).  Agent $i$ adopts the new technology if the
pay-off is positive ($P > 0$).  Otherwise it remains with the old
one. We remark that the pay-off of Eq.~\ref{eq:payoff} with $J_i>0$ is very
similar to the one used in a study of markets with
externalities~\cite{mirta1}. The difference resides in the fact that
in Ref.~\cite{mirta1} the positive quantity is the willingness to buy,
that changes from agent to agent, and the negative term is
the price that is the same for all agents. Here the positive term
(advertising) is constant and the resistance to adoption is negative
and idiosyncratic (changes from agent to agent).

The present model, condensed in the pay-off expression
Eq.(\ref{eq:payoff}), is a mean field model, meaning that every agent
``feels'' the influence of all the rest.  As it was stated previously,
we have added a second feature: the presence of a certain percentage
of agents with $J_i<0$ which act against the innovation, being the 
opposition stronger the higher the number of adopters. We denominate 
these agents, following Ref.~\cite{galam}, ``contrarians'', 
even when our definition is not exactly the same of Galam. 
He defines a contrarian as an agent that acts always against 
the majority, here, a contrarian agent,
when selected in the simulation, interacts with the adopting agents with
a negative $J_i$.
While Eq.~\ref{eq:payoff} for $J_i<0$ resembles an anti-ferromagnetic
interaction, it is not, because $J_i$ does not weigh a pair
interaction between two agents, but it weighs the interaction between one
agent and the bulk.

Another variant of the model considers groups of influence.  In this
case, each agent, instead of experiencing the effect of all the others, feels the
influence of a subset of them, selected at random from the population.

We also investigate a different distribution of the idiosyncratic
resistance to change $u_i$, considering a triangular distribution as
the simplest approach to the Gaussian distribution proposed in
Ref.~\cite{Rogers}.

In the next section we present the results for the basic model and the
three variants.

\section{Results}
\label{results}
A mean field analysis of the basic model (no contrarians, i.e.,
$J_i=J>0$ for all the agents, uniform distribution of $u_i$, and no
groups of influence) is readily straightforward.  We first describe
this case where some results can be obtained in an analytical way
and verified by the simulations.

\subsection{Basic model}
\label{model1}
\begin{figure*}[htp]
\centering \includegraphics[width=9cm,clip=true]{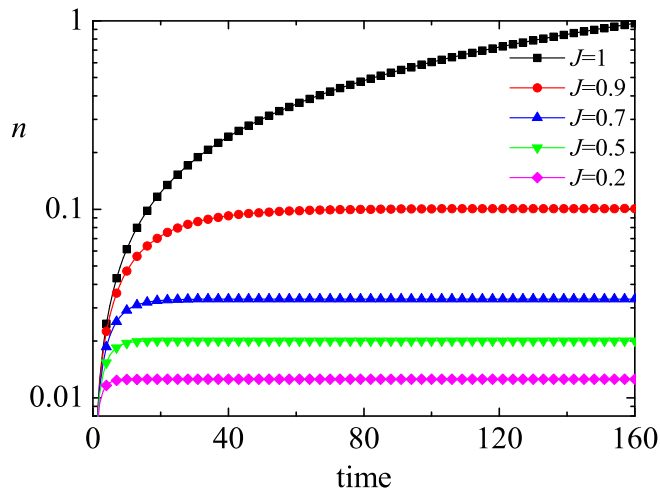}
\caption{(color online) Fraction of new adopters as a function of time
  (in MC steps) for marketing value $A=0.01$, system size $N=10^7$,
  and for different values of the coupling parameter $J$. The effect
  of the social interactions is remarkable. It is evident that it is
  not possible to have full adoption when $J < 1$.}\label{varJ}
\end{figure*}

\begin{figure*}[htp]
\centering \includegraphics[width=9cm,clip=true]{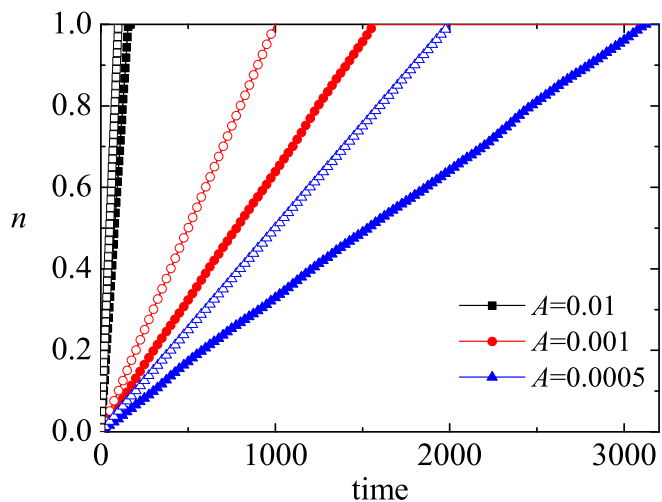}
    \caption{(color online) Fraction of new adopters as a function of
      time (in MC steps) for $J=1$, $N=10^7$ and different marketing
      values $A$. Each curve is a single run.  Even for a very small
      advertising ($A=0.001$) full adoption is obtained if $J=1$. Full
      symbols correspond to pick agents randomly, whereas open symbols
      correspond to select them sequentially.}
    \label{varA}
\end{figure*}

Starting with the whole population not having the new technology
($n=0$ in Eq.~\ref{eq:payoff}), the fraction of agents for which $u_i
\leq A$ will shift to the new technology. They will become the new
adopters, or the innovators according to Rogers~\cite{Rogers}, those
who can be convinced by the sole effect of publicity.  How many
individuals will adopt the new product as soon
as they are aware of it will depend on the force of the campaign
measured by $A$. As the values of $u_i$ are randomly and uniformly
distributed in the $[0,1]$ interval, the number of individuals with
values of $u_i$ below $A$ is $N_{adop} \approx A N$. Then, after the
first time step we have a fraction $n=N_{adop}/N=A$ of new adopters. In the
next step (of an arbitrary unit time) those who have $u_i \leq
(A+Jn)=(A + J A)$ will convert too, i.e. those who had less temptation
to try the novelty but will go after seeing that some people already
adopted it. If $J=1$, the adopters after the second step will be those
with $u_i \leq 2A $.  Following in this way, the society evolves, at
constant pace, into full adoption after a finite number of steps,
equal to $A^{-1}$. If $J <1$ however, the iterative process give rise
to a geometric series describing the dynamics: $n(t) = A \sum_{i=0}^t
J^i$, with asymptotic value $n(\infty) = A /(1-J)$. Thus, for $J=1$,
at any value of $A$, the whole society will adopt the new technology;
the intensity of $A$ will just determine the speed of this
transformation. For lower values of $J$ there will be a finite
fraction of the population adopting the innovation. These results are
verified by numerical simulations, as can be seen in
Fig.~\ref{varJ}. The simulation is performed using a Monte Carlo (MC)
method, i.e., choosing one agent at random and applying
Eq.~\ref{eq:payoff}. If the payoff is positive the agent become an
adopter. One MC step corresponds to choosing N agents at random, where
N is the size of the system.
We can see in the figure (Fig.~\ref{varJ}) some examples of the
temporal evolution of the fraction of new adopters for $N=10^7$, different
values of the coupling parameter $J$ and a fixed marketing value
$A=0.01$. The effect of the social interactions is remarkable. It is
evident that it is not possible to have full adoption when $J < 1$. On
the other hand, in Fig.~\ref{varA} we show a complementary point of
view: considering again $N=10^7$ agents, but for a fixed value of the
coupling parameter, $J=1$, we show the fraction of adopters as a
function of time for different values of the advertising $A$. We can
verify that the number of adopters increases linearly with time but
the slope is lower than $A$, the analytically predicted value. This is
so because when doing the MC simulation in a finite system, the
population is not covered exhaustively.  As a consequence, agents will
change opinion only when ``visited'' in the MC dynamics.
The probability of an agent not to be chosen in a MC
step is $(1-1/N)^N$, which in the limit of large $N$ is equal to
$\exp(-1)$; then the fraction of chosen agents is $1 - \exp(-1) =
0.632$, so the slope is $0.632A$ (as can be checked in Fig.~\ref{varA}). If the
simulations are performed sequentially, picking all the agents at each
MC step, the speed of the adoption is very close to $A$ (open symbols
in Fig.~\ref{varA}).  But even if the evolution is slower in the
non-sequential MC dynamics, the final values are not affected by the
type of simulation. Therefore, we decided to keep the MC type
simulation as we think it better represents the dynamics of a real
society.

The results for this kind of society are quite simple: if the
social interaction $J$ is strong enough, after a time that depends on
the intensity of the advertising, the full society will adopt the
innovation. On the other hand if the social influence is weak compared
with the maximum value of the idiosyncrasy $u_i$ of the agents, only a
finite fraction, which depends on the values of $J$ and $A$, will adopt
the innovation.

\subsection{Model with contrarians}
\label{model2}
\begin{figure*}[htp]
\centering \includegraphics[width=9cm, clip=true]{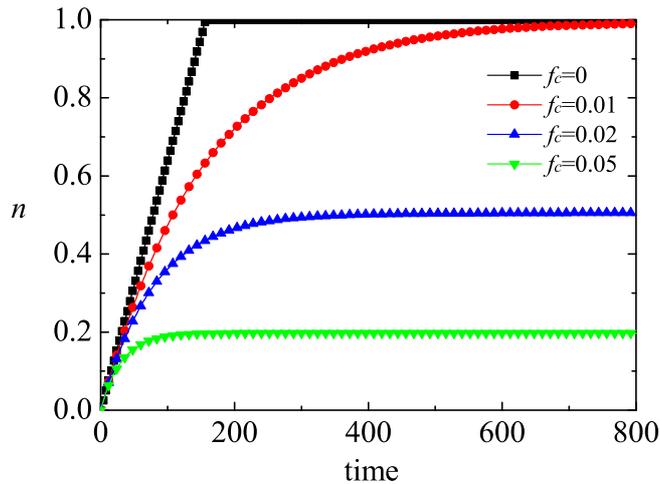}
\caption{(color online) Fraction of adopters of the new technology as
  a function of time, for four different values of the fraction of
  contrarians: $f_c=0$, $0.01$, $0.02$ and $0.05$.  In all cases,
  $A=0.01$ and $N=10^7$.}\label{contra}
\end{figure*}

\begin{figure*}[htp] \centering \includegraphics[width=9cm,clip=true]
{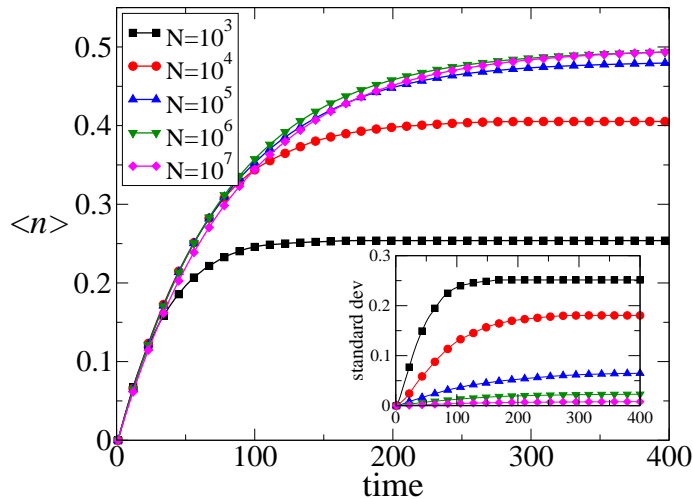}
\caption{(color online) Finite size effects: Average time evolution of
  the system for different sizes. For each curve, symbols represent
  averages taken over $100$ runs. Inset: The standard deviation of the
  averages shown in the main panel. For $N \geqslant 10^6$ we can say
  that the evolution is independent of the size of the
  sample.} \label{size}
\end{figure*}

\begin{figure*}[htp] \centering \includegraphics[width=12cm, clip=true]
{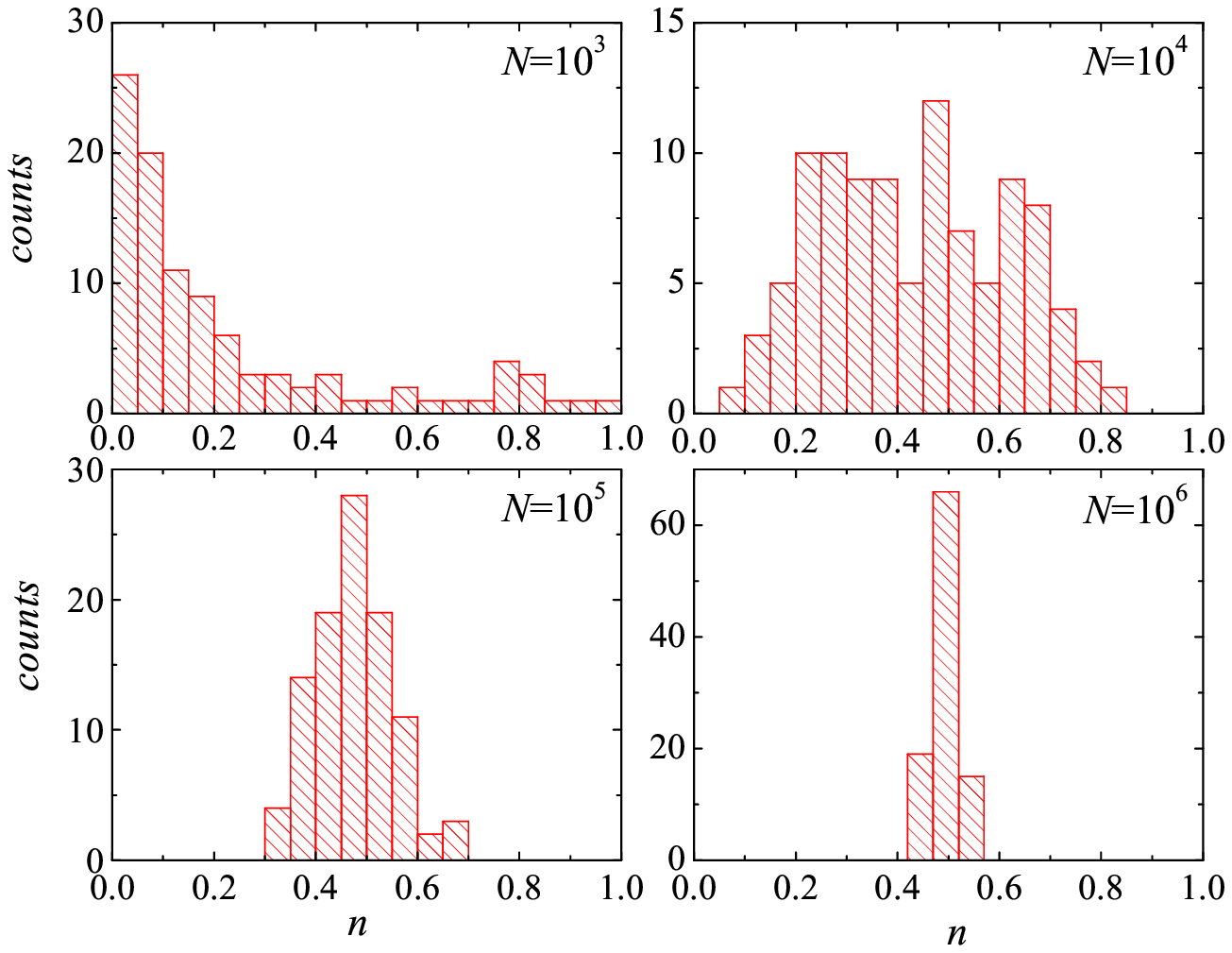}
\caption{Finite size effects: Histograms of the final number of
  adopters for $100$ samples and four different system sizes from
  $10^3$ to $10^6$, with $A=0.01$ and a fraction of contrarians
  $f_c=0.02$.}
 \label{histo}
\end{figure*}

In this section we explore the effect of having a part of the
population reacting in opposition to the majority.  We call these
agents contrarians and they are characterized, in terms of the model,
by having a negative $J_i$ coupling.  All along this section we assume
$|J_i|=1$ and, as agents do not change their idiosyncrasy in time, the
number of contrarians will be constant in time. We define the fraction
of contrarians, $f_c$, as the portion of the population having $J_i=-1$.
We have represented in Fig.~\ref{contra} the fraction of agents
adopting the new technology as a function of time for a fixed value of
the advertising, $A=0.01$, and four different fractions of
contrarians. The size of the system is $N=10^7$ agents. In the
absence of contrarians we recover the previous result: the entire
society adopts the innovation. Yet, a small concentration of
contrarians is enough to reduce the fraction of adopters by a
significant fraction; for example $f_c=0.02$ decreases the fraction of
adopters by a half.
We must say, however, that these results strongly depend on the
size of the samples.  On Fig.~\ref{size} we represent an average over $100$
samples of the temporal evolution for different system sizes, for
the case $A=0.01$, and $f_c=0.02$. While in the previous figure
(Fig.~\ref{contra}, $N=10^7$) the fraction of adopters converge to $50\%$ for
$f_c=0.02$, we can see that for smaller samples the number of adopters
may be bigger, up to $75\%$ for $N=10^3$. Also, the fraction of
adopters exhibit large fluctuations for small systems at all times, as
it can also be observed on Fig.~\ref{size}. The dispersion goes down
with $1/\sqrt{N}$.  The finite size effects can also be visualized by
performing an statistical study over the final state for the different
samples. On Fig.~\ref{histo} we have represented by histograms the
distribution of the asymptotic number of adopters over $100$ samples
and four different sizes of the system. We can verify that for smaller
systems the distribution is very wide ---even Poisson-like for
$N=10^3$ agents---, but they became narrower as the system size
increases.  After these considerations, in the following (and as
it was in the previous section) we consider a number of agents
$N=10^7$.

\begin{figure*}[htp] \centering \includegraphics[width=9cm,
clip=true]{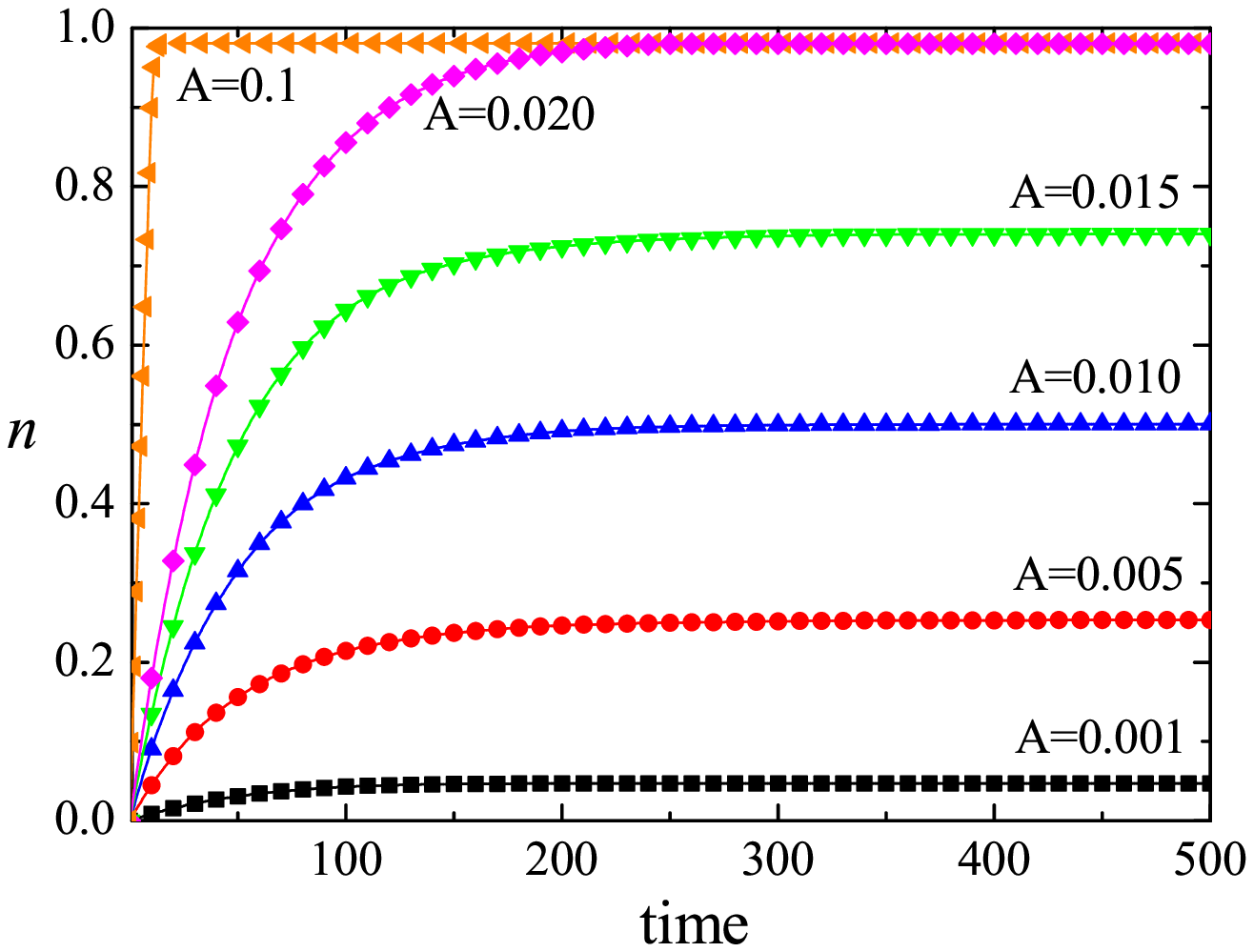} \caption{(Color online) Effect of the
    advertising $A$ for a fixed percentage of contrarians
    $f_c=0.02$. Each curve is a single run with $N=10^7$.}
\label{adver}
\end{figure*}

\begin{figure*}[htp] \centering \includegraphics[width=9cm,
clip=true]{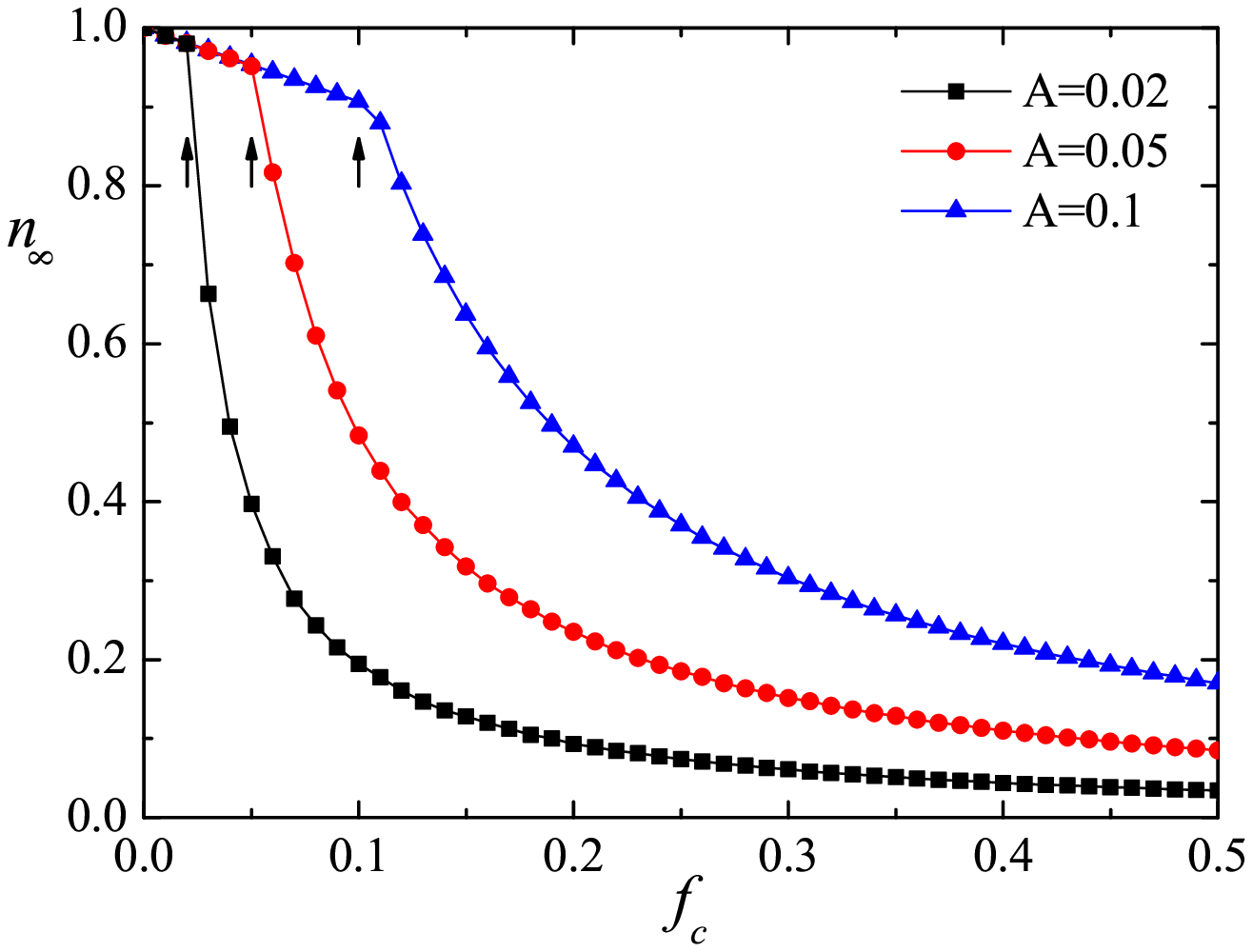} \caption{(Color online) Final proportion of
    adopters as a function of the fraction of contrarians for three
    different values of the advertising. Each point represents the
    final state of a single run with $N=10^7$ (lines are just a guide
    to the eye). Arrows indicate the critical value $f_c=A$.}
\label{cont_adv}
\end{figure*}

Now we turn to analyze the simultaneous effect of advertising and
contrarians. This is represented in Fig.~\ref{adver} for a fixed
percentage of contrarians ($f_c=0.02$) and different values of $A$. It
is clear from the figure that there is a critical value of advertising
to obtain the maximum adoption of the innovation by the society. Due
to the presence of contrarians the maximum fraction of adopters is
bounded by $1-f_c$.  The number of adopters increases linearly as the
number of contrarians decreases or as the advertising increases, until
$f_c=A$, when a clear change in the behavior is
observed. Fig.~\ref{adver} shows that when $A \geq f_c$ the maximum
possible adoption ($1-f_c$) is reached. This can be best appreciated by
looking at it from another angle: for a fixed advertising value $A$,
above the critical value $f_c=A$ the number of adopter falls abruptly
and in inverse proportion to $f_c$, as shown in
Fig.~\ref{cont_adv}. Interestingly, a relatively small percentage of
contrarians may block the maximum possible adoption of the innovation
or force a stronger advertising (i.e. $A \ge f_c$) campaign in order
to impose the innovation.  Therefore, it is evident from
Figs.~\ref{adver} and \ref{cont_adv} that the system converges to the
maximum adoption only when the advertising is strong enough, $A \ge
f_c$, and also that the number of adopters grows with time, more and
less linearly, up to a saturation point.

\begin{figure*}[htp] \centering \includegraphics[width=9cm, clip=true,
angle=-90]{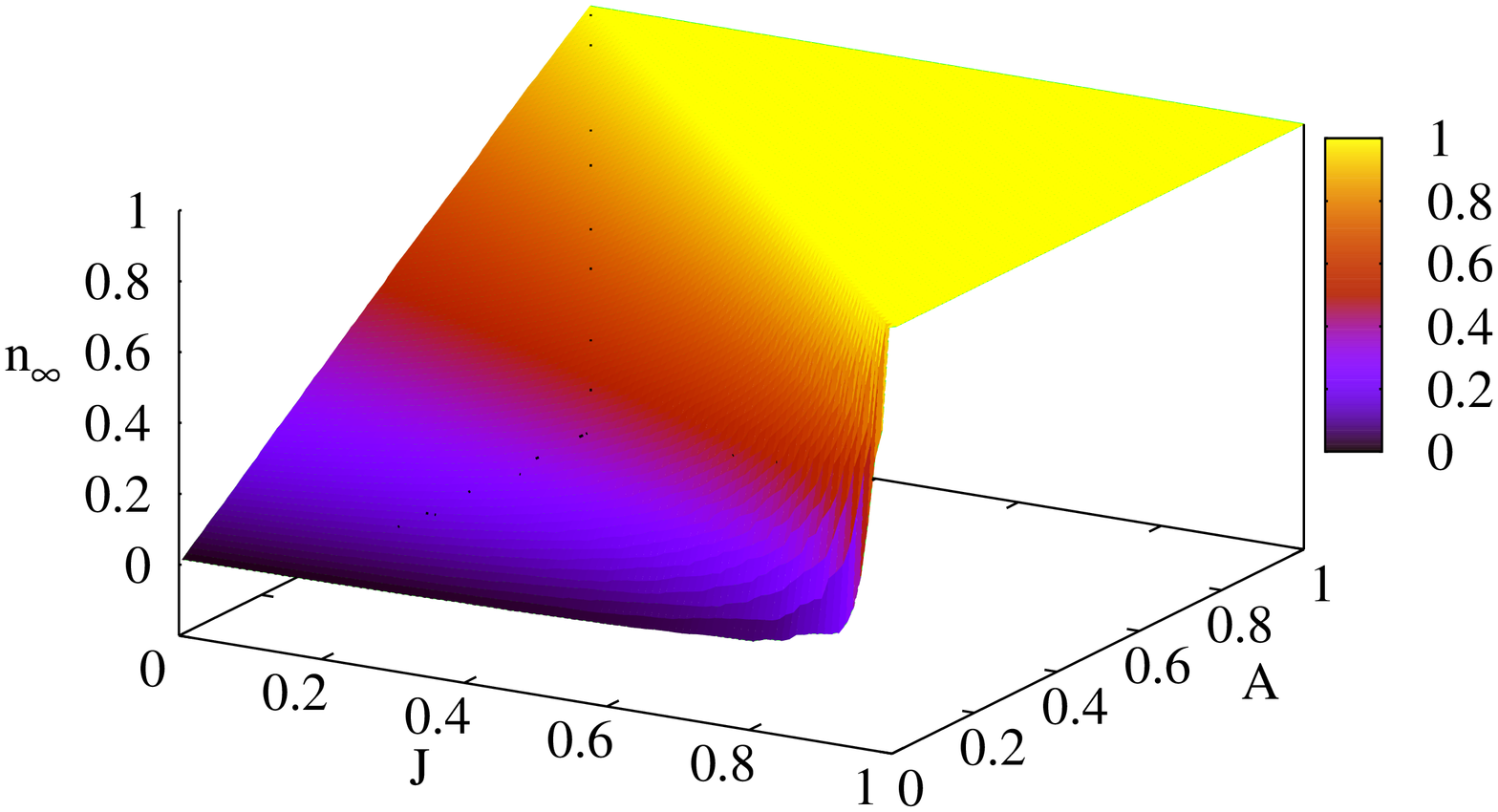}
   \caption{(Color online) Final fraction of adopters as a function
     of $J$ and $A$, for $f_c=0.02$}
    \label{diag}
\end{figure*}

The results of the first two subsections can be summarized in a $3D$
diagram. On Fig.~\ref{diag} we show the final fraction of adopters on
the $z$-axis as a function of the social interaction $J$ on the
$x$-axis and of the advertising $A$ on the $y$-axis, for a fixed
fraction of contrarians: $f_c=0.02$. A sharp but
continuous transition as a function of $A$ is observed for high values of $J$,
while the transition is smoother and converges to a smaller number of
adopters for intermediate values of $J$ and $A$.

Finally, we would like to emphasize that allowing or not the
contrarians to adopt the new technology is mostly irrelevant. In other
words, allow them to check their pay-off function
(Eq.~\ref{eq:payoff}) to take a decision, or keep them in a permanent
non-adopter state, yields the same result. This is because the higher
the number of adopters $n$, the stronger the {\it negative} effect on
the payoff of the contrarians, and thus it will be very difficult for a
whimsical contrarian to become an adopter, even if her/his resistance
$u_i$ is low.
In effect, our contrarians are much closer to inflexibles than to the 
contrarians of Galam~\cite{galam2011b}.

\subsection{Groups of influence}
\label{groups}
In the preceding section we have implicitly considered that each agent
in the society knows exactly what all the other agents have chosen, because
we used $n=N_{adop}/N$ (where $N_{adop}$ is the number of adopters of
the new technology and $N$ is the total number of agents) as the
fraction of adopters to plug into its payoff.  However, nobody
can have such instant and exact account of how the rest
of the society (or country, or city) is reacting to the novelty. Most
probably he/she will have an imperfect sense of that through the news,
or better from what people around him/her (family, friends,
colleagues, etc) are doing.  Besides, from the technical point of view, we have seen
that the size of the sample is important for the results. Then, we
consider in this subsection that people interact with a group much
smaller than the whole society, i.e. the number of members of the
influence group is $N_g << N$. We calculate the payoff using the same
Eq.~\ref{eq:payoff} but now with $n'=N_{adop,g}/N_g$, where
$N_{adop,g}$ is the number of adopters within the group of influence
of size $N_g$.

\begin{figure*}[htp] \centering \includegraphics[width=9cm, clip=true]
{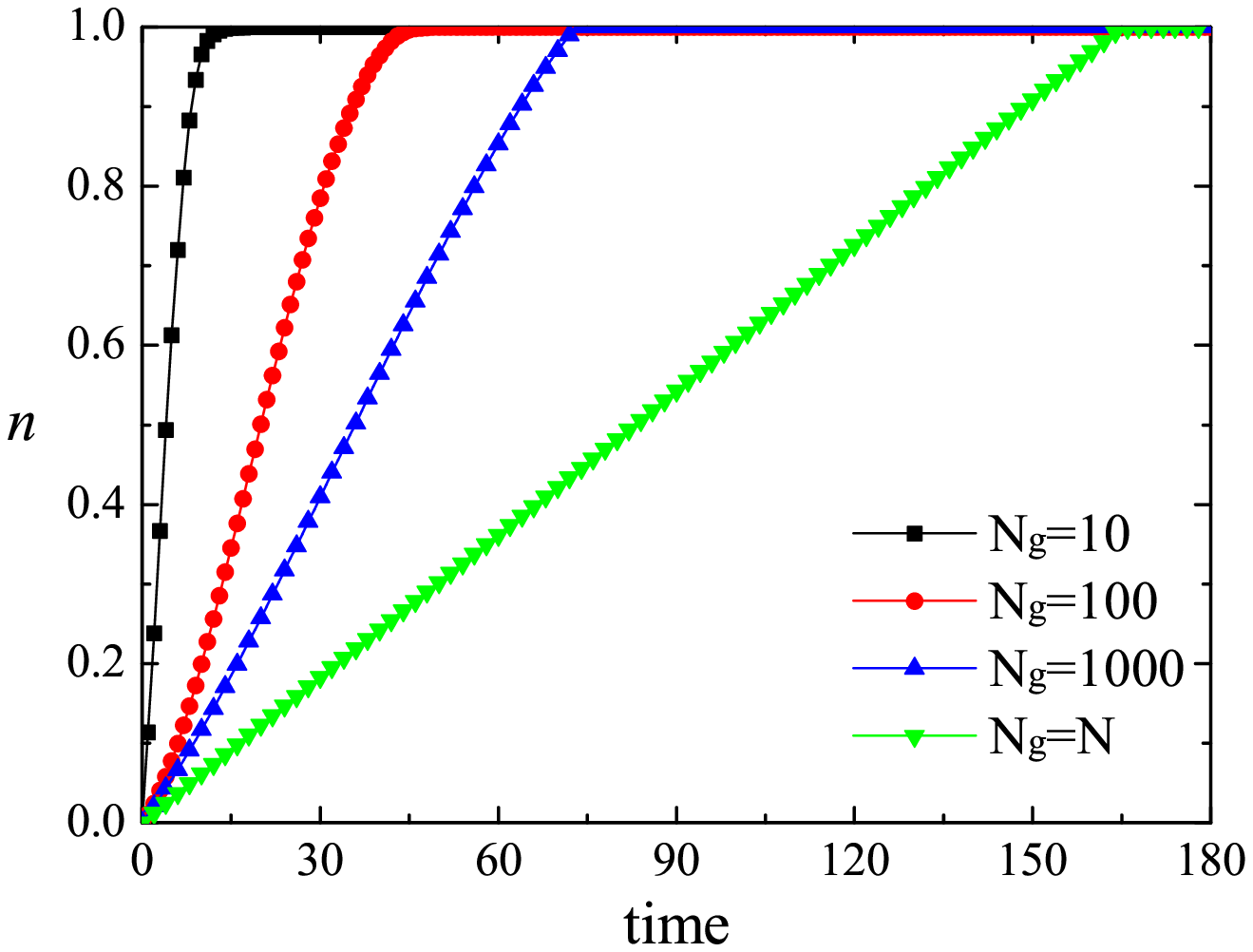}
    \caption{(Color online) Number of adopters for $N=10^7$, $A=0.01$,
      $J=1$, and zero contrarians, with three different sizes of the
      groups of influence: $N_g=10$, $100$ and $1000$. It is worth to
      note that the convergence to a state of complete adoption is faster
      the smaller is the group of influence. This means that close
      contacts are more effective than the ``diluted'' interaction
      with all the agents.}
\label{influ}
\end{figure*}

While in the real life the groups of influence can vary from person to
person, we make a simpler approach here assuming that all the groups of
influence have the same size, and that the members of the groups are
chosen at random.
In other words, at each MC step and for each agent, we choose a set of
$N_g$ other agents at random from the population and the state of them
is used in the payoff. So the size of the groups is fixed and the same
for all agents, while the members of them are chosen at random at every
time step.  Notice that groups are not isolated because any group has
overlapping members with another group (in principle it is possible but
highly improbable for a group to be isolated).
We first analyze the case without contrarians. We
have represented on Fig.~\ref{influ} the time evolution of the number
of adopters for similar parameters than in the previous section:
$J=1$, $A=0.01$, $N=10^7$ and for different sizes of the groups of
influence. It can be observed that the system converges to full
adoption faster when the size of the groups of influence is
smaller. This means that the effect of small groups is more effective
to spread the new technology. Comparing this results with the case
with zero contrarians, represented in Fig.~\ref{contra}, is clear that
the time necessary to converge to complete adoption is much shorter
with than without groups of influence. Moreover, we have verified that
in the case of groups of influence the determinant factor is the size
of the groups and not the size of the whole society. In other words,
the evolution of the number of adopters for a fixed size of the group
of influence is almost independent of the sample size.

\begin{figure*}[htp] \centering \includegraphics[width=6.5cm,
clip=true ]{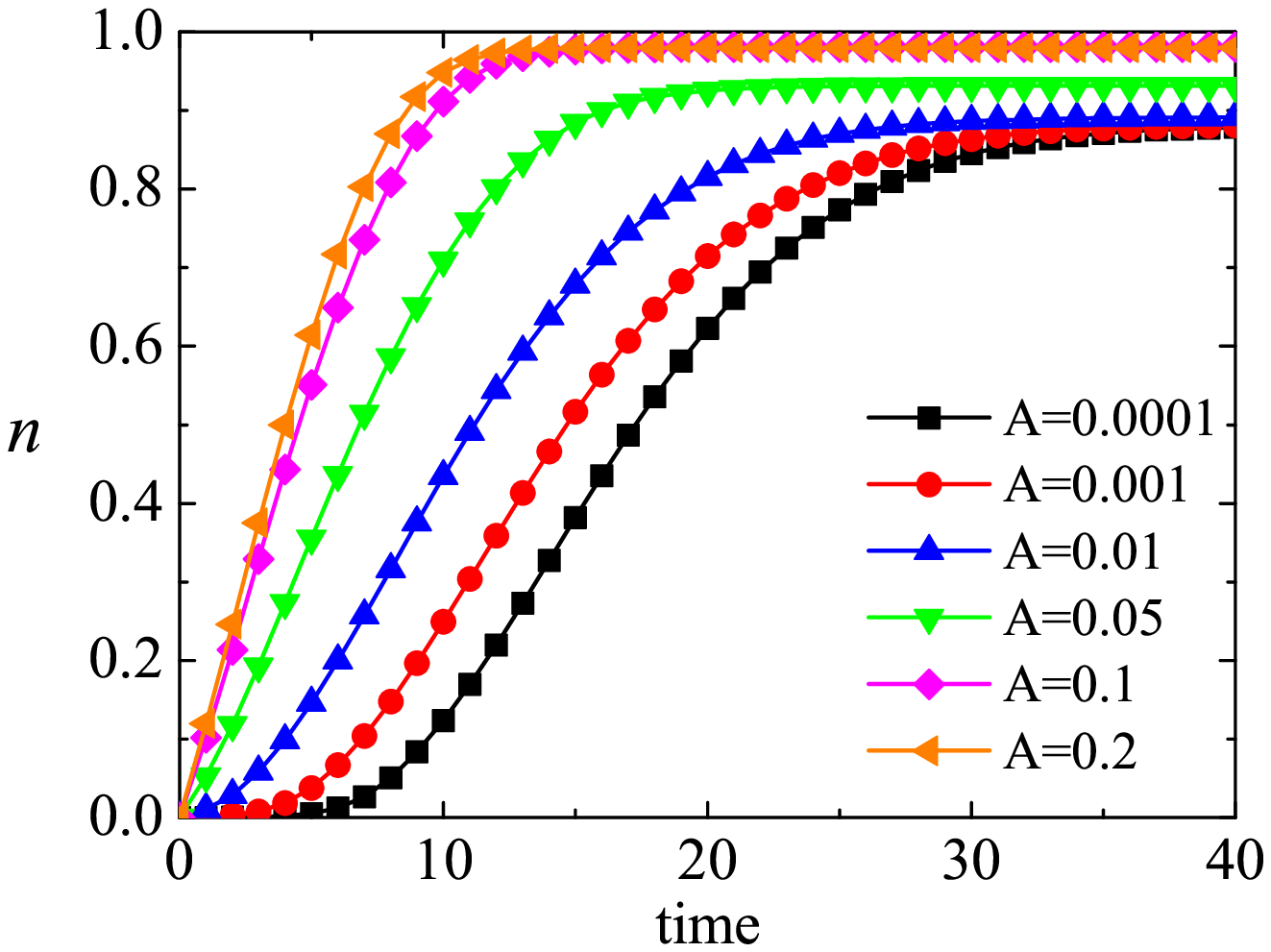} \includegraphics[width=6.5cm,
clip=true ]{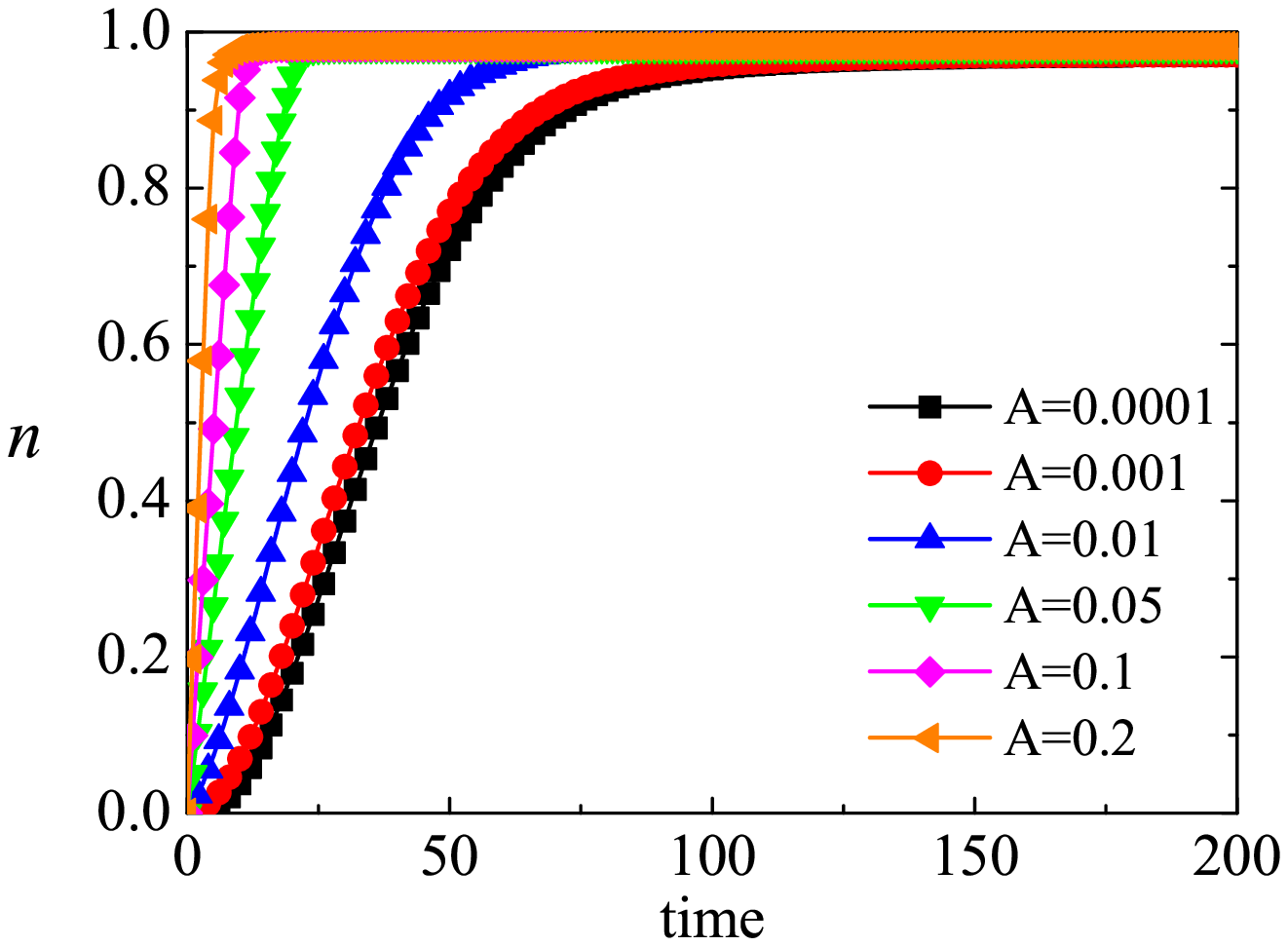}
\caption{(Color online) Evolution of the number of adopters for $J=1$,
  $f_c=0.02$, and $N=10^7$, for different values of the advertising $A$
  and for two groups of influence: $N_g=10$ (left panel) and $N_g=100$
  (right panel). Notice that the time scale is different in the
  two panels, although the final fraction of adopters is, in all the
  cases, $n=1-f_c=0.98$.}
\label{influ_contra}
\end{figure*}

If we include now the contrarians, the results are very rich and in
some cases correspond very well to the adoption empirical curves of US
consumers, represented in Fig.\ref{Gauss}. To study in full detail
the combined effect of contrarians and groups of influence we have
represented different cases in Fig.~\ref{influ_contra}. One evident
result is that the number of adopters increases as swiftly as in the
case without contrarians. However, and at variance with previous
results, even for $A \geq f_c$ the system does not arrives at full
adoption, i.e. the effect of the contrarians is somehow amplified by
the groups of influence: the smaller the group, the bigger the
effect. To verify this result, in the left panel of
Fig.~\ref{influ_contra} we have represented the results for a group of
influence $N_g=10$, while in the right panel we show the results for
$N_g=100$. When increasing the size of the group of influence, other
parameters fixed, the system approaches a behavior closer to the mean
field results.

In short, the effect of groups of influence is very interesting and
induces behaviors that are different from the mean field results, and
closer to practical situations. The importance of the effect of
interactions in small groups in driven the society to adopt a given
opinion or decision has been extensively discussed by
Galam~\cite{galam,galam2011} and similar results are also reported in
ref.~\cite{Hub} when comparing the effect of friends and global opinions
when taking a decision to buy or not a product. In any case, it is worth
notice that the adoption curves, when groups of influence are included,
look more similar to the empirical results quoted in the
introduction~\cite{Rogers,Forbes,NYT}.

\subsection{Triangular Distribution}
\label{triang-distr}
A possible objection to the preceding results is that we considered a
uniform distribution of the idiosyncratic resistance to change $u_i$,
while the hypothesis of Rogers~\cite{Rogers} is that this distribution
is Gaussian (See Fig.~\ref{Gauss}). We can argue that the results with
groups of influence reproduce in a qualitative way the proposed curve
of adoption of Rogers, and also the empirical
curves~\cite{Forbes,NYT}, but for the seek of completeness we
investigate a different distribution of the idiosyncratic resistance
to change. Let us briefly consider a triangular distribution centered
in $\langle u \rangle=0.5$, that is a rough but simple approach to the
Gaussian distribution.

\begin{figure*}[htp] \centering \includegraphics[width=6.5cm,
clip=true ]{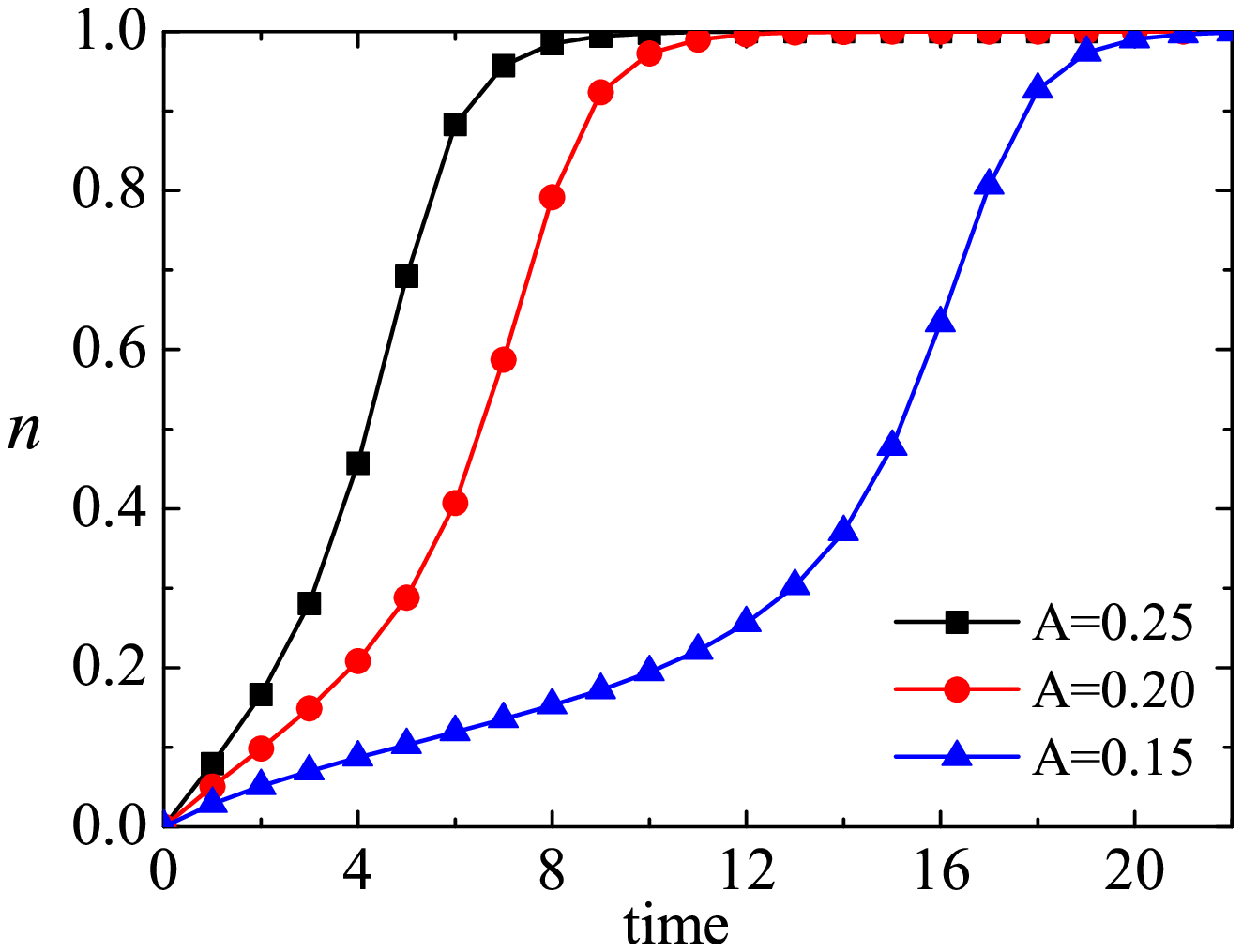}\includegraphics[width=7cm,
    clip=true ]{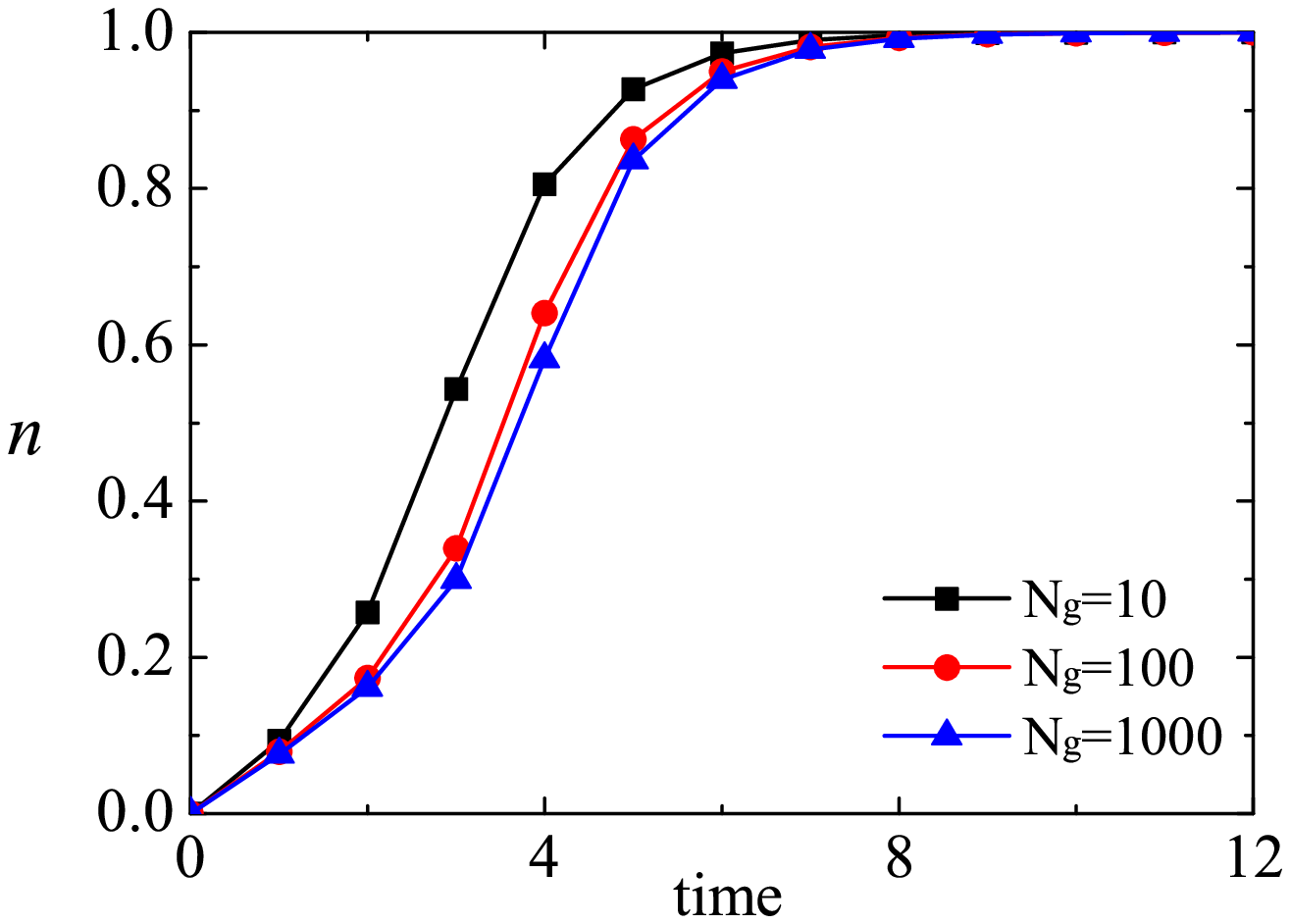}
\caption{(Color online) Evolution of the number of adopters for a
  triangular distribution of idiosyncratic resistance to change and
  for $J=1$, $N=10^7$, and no contrarians. Left panel: different
  values of the advertising $A$ without groups of influence. Right
  panel: different sizes of the groups of influence for $A=0.2$.}
\label{influ_triang}
\end{figure*}

On Fig.~\ref{influ_triang} we have represented the adoption curves for
a triangular distribution of $u_i$ and without contrarians. The left
panel shows the results for the mean field model (i.e., the model with
no groups of influence) and three different values of the advertising
$A$. In the right panel, the effect of different sizes for the groups
of influence is displayed.  It is worth to say that, in order to have
the expected behavior, we have to consider higher values of the
advertising. This is because the triangular distribution has a smaller
number of early adopters when compared with the uniform
distribution. We have also verified that with the triangular
distribution of $u_i$, contrarians have a very weak effect, so
results considering contrarians have been omitted.

\begin{figure*}[htp]
\centering \includegraphics[width=9cm, clip=true,
  angle=-90]{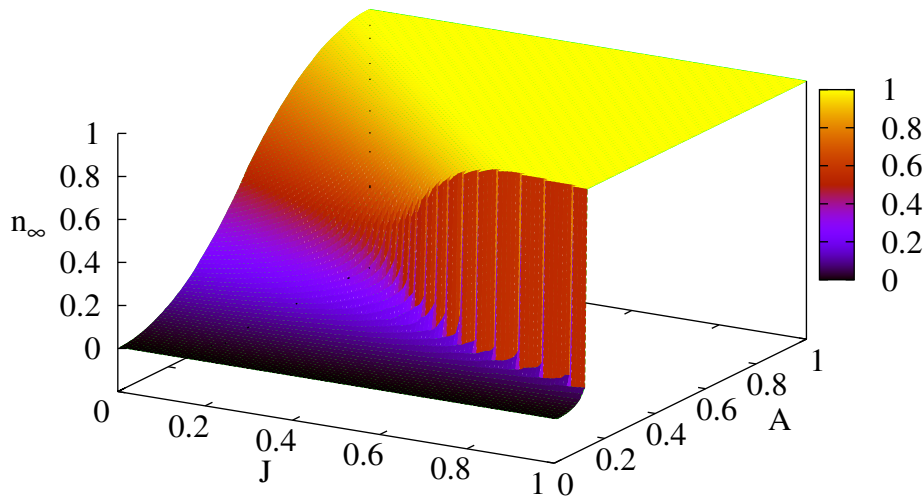}
    \caption{(Color online) Phase diagram of the final fraction of new
      adopters $n$ as a function of the coupling parameter $J$ and of
      the advertising $A$.}
    \label{triang}
\end{figure*}

Besides, the inclusion of a non-uniform (triangular) distribution of
$u_i$ brings a novel result which is showed in Fig.~\ref{triang}.  Such
figure represents, in $3D$, the asymptotic fraction of adopters,
$n$, as a function of $J$ and $A$, exhibiting a clear
phase transition as a function of $A$ for values of $J \ge 0.5$. This
is in agreement with the phase transition obtained in
Ref.~\cite{mirta1} and it means that the number of adopters can suffer
an abrupt and discontinuous change as a function of the advertising.

\section{Discussion and Conclusions}
We have presented a model of adoption of a new technology that, in
spite of its simplicity, captures the essential features of the
behavior of societies: a distribution of an idiosyncratic resistance
to change, the power of advertising, the effect of social
interactions, and the presence of people who act against the general
behavior.  The effect of each one of these factors has been
investigated separately as well as in combination.

A relatively weak social interaction (either because of a low value of
the interaction parameter $J$ or by a low advertising $A$), can impede
the adoption of the new technology by the full society. Such can be
the case of products with a limited penetration like the dishwasher or
the blue-ray player.

Moreover, the inclusion of a small concentration of contrarians in the
model is enough to reduce the fraction of adopters by a significant
fraction. They may force a stronger advertising campaign in order to
impose the innovation.

Regarding again the social interaction, we observed that small groups
of influence may be a determinant factor both in the speed of adoption
as well as in the percentage of the population adopting the new
technology. We have verified that the interaction with small groups
instead of with the full society may explain the general behavior of
the empirical adoption curves~\cite{Forbes,NYT}.  Also, different
speeds of adoption may be obtained by changing the size of the groups
of influence, the strength of the advertising, or the percentage of
contrarians.

Considering a triangular distribution of the idiosyncratic resistance
to change can also result in adoption curves with profiles similar to
the empirical results. Besides, in this case it can exists an abrupt
transition in the final number of adopters as a function of the
advertising. Otherwise, even if a relatively slow growth of adoption
may be observed in some cases in Fig~\ref{influ_triang}, the
triangular distribution does not provide a wide variety of speeds of
adoption. In other words, the triangular distribution seems to be
insensitive to the size of the groups of influence.

Our results indicate that, on one side, the acceleration of the
adoption velocity is a combined effect of the advertising
(Fig.~\ref{varJ}) and of the size of the groups of influence
(Fig.~\ref{influ_contra}). On the other side, attaining or not full
adoption depends on both, the strength of the social interaction and
the fraction of contrarians.

Finally we want to emphasize that the present model represents a mean
field approach, regarding the interaction with the bulk (or a random
subgroup), without taking into account any possible network topology.
Remarkably, in spite of that, the model is able to reproduce the
essential features of the adoption of a new technology.  However, we
realize that the effect of distributing the agents on a network could
have some relevant effect in the dynamics of technology adoption, and
such is the subject of a work already in progress.

\section{Acknowledgments}
JRI thanks the kind hospitality of the Statistical and
Interdisciplinary Physics Group of Instituto Balseiro in S.C. de
Bariloche, Argentina and acknowledges the financial support of a
``Dr. C\'esar Milstein'' grant from the program Ra\'{\i}ces of the
Ministerio de Ciencia Tecnolog\'{\i}a e Innovaci\'on Productiva of
Argentina.

The authors acknowledge support from Brazilian agency CNPq and
from projects CNPq/Prosul 49044/2007 and FAPERGS-Pronex.

\end{document}